\documentclass[aps,pra,twocolumn,showpacs,floatfix]{revtex4-1}

\setcitestyle{square}
\usepackage{epsfig,amsmath,amssymb}
\usepackage{graphics} 
\usepackage{subfigure}
\usepackage{color}
\usepackage{multirow}
\usepackage{mathrsfs}
\usepackage{natbib}

\begin{document}

\title
{
Quasiclassical magnetic order and its loss in a spin-$\frac{1}{2}$ Heisenberg antiferromagnet on a triangular lattice with competing bonds
}

\author
{P.~H.~Y.~Li}
\email{peggyhyli@gmail.com}
\author
{R.~F.~Bishop}
\email{raymond.bishop@manchester.ac.uk}
\affiliation
{School of Physics and Astronomy, Schuster Building, The University of Manchester, Manchester, M13 9PL, UK}

\author
{C.~E.~Campbell}
\affiliation
{School of Physics and Astronomy, University of Minnesota, 116 Church Street SE, Minneapolis, Minnesota 55455, USA}

\begin{abstract}
  We use the coupled cluster method (CCM) to study the
  zero-temperature ground-state (GS) properties of a
  spin-$\frac{1}{2}$ $J_{1}$--$J_{2}$ Heisenberg
  antiferromagnet on a triangular lattice with competing nearest-neighbor and
  next-nearest-neighbor exchange couplings $J_{1}>0$ and $J_{2} \equiv
  \kappa J_{1}>0$, respectively, in the window $0 \leq \kappa < 1$.  The
  classical version of the model has a single GS phase transition at
  $\kappa^{{\rm cl}}=\frac{1}{8}$ in this window from a phase with
  3-sublattice antiferromagnetic (AFM) 120$^{\circ}$ N\'{e}el order
  for $\kappa < \kappa^{{\rm cl}}$ to an infinitely degenerate family
  of 4-sublattice AFM N\'{e}el phases for $\kappa > \kappa^{{\rm
      cl}}$.  This classical accidental degeneracy is lifted by quantum
  fluctuations, which favor a 2-sublattice AFM striped phase.  For the
  quantum model we work directly in the
  thermodynamic limit of an infinite number of spins, with no
  consequent need for any finite-size scaling analysis of our results.
  We perform high-order CCM calculations within a well-controlled
  hierarchy of approximations, which we show how to extrapolate to
  the exact limit.  In this way we find results for the case $\kappa =
  0$ of the spin-$\frac{1}{2}$ model for the GS energy per spin,
  $E/N=-0.5521(2)J_{1}$, and the GS magnetic order parameter,
  $M=0.198(5)$ (in units where the classical value is $M^{{\rm
      cl}}=\frac{1}{2}$), which are among the best available.  For the
  spin-$\frac{1}{2}$ $J_{1}$--$J_{2}$ model we find that the classical
  transition at $\kappa=\kappa^{{\rm cl}}$ is split into two quantum
  phase transition at $\kappa^{c}_{1}=0.060(10)$ and
  $\kappa^{c}_{2}=0.165(5)$.  The two quasiclassical AFM states (viz.,
  the 120$^{\circ}$ N\'{e}el state and the striped state) are found to
  be the stable GS phases in the regime $\kappa < \kappa^{c}_{1}$ and
  $\kappa > \kappa^{c}_{2}$, respectively, while in the intermediate
  regimes $\kappa^{c}_{1} < \kappa < \kappa^{c}_{2}$ the stable GS
  phase has no evident long-range magnetic order.
\end{abstract}


\maketitle

\section{INTRODUCTION}
\label{introd_sec}
Quantum Heisenberg antiferromagnets (HAFMs), comprising spins (with
spin quantum number $s$) on an infinite regular lattice in two spatial
dimensions, and interacting via a pure nearest-neighbor (NN)
Heisenberg potential with exchange coupling $J_{1}>0$, have long
occupied a special role in the theory of quantum phase transitions.
Thus, for example, the well-known Mermin-Wagner theorem
\cite{Mermin:1966} proves that in both one and two dimensions HAFMs
are disordered at any nonzero temperature ($T \neq 0$), in the sense that
thermal fluctuations completely destroy all long-range order (LRO).
Similarly, in one dimension quantum fluctuations destroy the N\'{e}el
LRO even at zero temperature ($T=0$).  On the other hand, the
Mermin-Wagner theorem does {\it not} prohibit a ground state (GS) with
LRO for any two-dimensional (2D) model with a continuous symmetry.

It thus remains an open question as to whether a particular 2D
spin-lattice model will or will not display LRO in its GS at $T=0$.
For a pure 2D HAFM both quantum fluctuations and any geometrical
frustration present in the lattice can potentially combine to destroy
long-range N\'{e}el-type order.  Quantum fluctuations are generally
larger for smaller values of $s$, stronger frustration, and lower
coordination number $z$.  Of the 11 uniform Archimedean lattices,
those tilings with the greatest frustration are the triangular lattice
(with $z=6$) and the kagome lattice (with $z=4$).  Thus, the
spin-$\frac{1}{2}$ HAFMs on the triangular and kagome lattices have
attracted much specific interest in the past.

For the triangular-lattice HAFM the classical ($s \rightarrow \infty$)
GS is a 3-sublattice N\'{e}el state with an angle of 120$^{\circ}$
between the spins on different sublattices, which thus breaks the
translational symmetry of the lattice.  Historically, some 40 years
ago, the spin-$\frac{1}{2}$ HAFM on the triangular lattice was the
first model to be proposed \cite{Anderson:1973_QSL,Fazekas:1972_QSL}
as a microscopic realization of a system whose GS might be a quantum
spin liquid (QSL).  It was argued that the GS might be similar to that
of the 1D HAFM, and it was thus proposed that it had the form of a
rotationally-invariant, resonating valence bond (RVB) state, instead
of a quasiclassical N\'{e}el state akin to the exact classical GS,
albeit with a reduced (but nonzero) value of the corresponding
sublattice magnetic order parameter.

By contrast, spin-wave theory (SWT)
\cite{Oguchi:1983_triang_SWT,Nishimori:1985_triang_SWT,Jolicoeur:1989_triang_SWT,Miuyaki:1992_triang_SWT,Chubukov:1994_triang_SWT,Chernyshev:2009_triang_SWT}
results, even at higher orders consistently predict that quantum
fluctuations on the spin-$\frac{1}{2}$ triangular-lattice HAFM do not
destroy the 120$^{\circ}$ N\'{e}el antiferromagnetic (AFM) LRO, but
lead to a reduction in the sublattice magnetization of around 50\%
from the classical value.  A number of variational calculations have
also been performed for the spin-$\frac{1}{2}$ triangular-lattice HAFM
with conflicting results.  While some calculations
\cite{Huse:1988_triang_VMC,Sindzingre:1994_triang_VMC} predict a
quasiclassical ordered state, others
\cite{Oguchi:1986_triang_VMC,Kalmeyer:1987_triang_VMC,Yang:1993_triang_VMC}
predict a magnetically disordered state.  Typically, however, the
former are based on variational wave functions with LRO built in from
the outset, while the latter typically employ a spin-liquid type of
wave function.

While many of the early numerical studies
\cite{Fujiki:1987_triang_ED,Fujiki:1987b_triang_ED,Nishimori:1988_triang_ED,*Nishimori:1989_triang_ED,*Nishimori:1989b_triang_ED,Imada:1987_triang_ED,*Imada:1989_triang_ED,Leung:1993_triang_ED}
for the spin-$\frac{1}{2}$ triangular-lattice HAFM based on the exact
diagonalization (ED) of small lattice clusters predicted a GS with no,
or very small, magnetic LRO, it was later pointed out rather
forcefully \cite{Bernu:1994_triang_ED} that two basic requirements
need to be carefully met in order to analyze the raw numerical ED data
properly.  Firstly, a consistent finite-size scaling analysis is
needed to reach the thermodynamic limit ($N \rightarrow \infty$) of a
lattice with $N$ spins, and secondly, a proper quantum definition of
observables needs to be made.  Bernu {\it et al}.\
\cite{Bernu:1994_triang_ED} argued that once those two constraints are
met, the numerical data point to an ordered ground state for the
infinite lattice.  It is clear that the $N \rightarrow \infty$
extrapolation is rather difficult for this model.  In a separate paper
\cite{Bernu:1992_triang_ED} Bernu {\it et al}.\ quote an extrapolated
ED value for the magnetic order parameter $M$ of approximately 50\% of
the classical value with a large error, probably of the order of $\pm
5\%$ or more.  This is in reasonable agreement with the corresponding
predictions of $M=47.74\%$ and $49.95\%$ of the classical value from
leading-order SWT and second-order SWT, respectively
\cite{Chernyshev:2009_triang_SWT}, in which $M$ is correct to order
$O(1/s)$ and $O(1/s^{2})$ respectively in the usual SWT $1/s$
expansion.  A more recent ED analysis \cite{Richter:2004_triang_ED}
quotes a more accurate, reduced value for $M$ of $38.6\%$ of the
classical value.

Series expansion (SE) methods constructed around an ordered state have
given equally confusing results for the spin-$\frac{1}{2}$ triangular
lattice HAFM.  For example, an early $T=0$ SE calculation
\cite{Singh:1992_triang_ED} found some evidence that this model may be
very close to a quantum critical point (QCP).  If ordered at all, the
model was estimated to have a value for $M$ much smaller than the SWT
estimates, and rather close to zero.  By contrast, a much more recent
SE study \cite{Zheng:2006_triang_ED} quoted a value for $M$ of $(38
\pm 4)\%$ of the classical value.

We turn finally to other recent calculations for the
spin-$\frac{1}{2}$ triangular-lattice HAFM, employing state-of-art
tools of microscopic quantum many-body theory.  Firstly, sequences of
clusters, using pinning fields and cylindrical boundary conditions to
provide for rapidly converging finite-size scaling ($N \rightarrow
\infty$), have been studied using the density-matrix renormalization
group (DMRG) method \cite{White:2007_triang_DMRG}.  Nevertheless, it was found
that the finite-size analysis for the triangular lattice HAFM was much
less accurate than that for the corresponding HAFM on the square
lattice, for example.  The best result thus obtained for the magnetic
order parameter of the spin-$\frac{1}{2}$ triangular lattice HAFM is
$M \approx (41 \pm 3) \%$ of the classical value.  Secondly, the
spin-$\frac{1}{2}$ triangular lattice HAFM has also been studied on
clusters of up to $N=144$ sites using the Green function Monte Carlo
(GFMC) method \cite{Capriotti:1999_trian}, together with a stochastic
reconfiguration technique that allows the fixed-node approximation
(which is needed to overcome the well-known minus-sign problem) to be
released in a controlled, albeit approximate, way.  The best estimate
thus obtained in the thermodynamic limit ($N \rightarrow \infty$) for
the magnetic order parameter of the spin-$\frac{1}{2}$ triangular
lattice HAFM is $M \approx (41 \pm 2)\%$ of the classical value.  It
is perhaps worth noting in this context that an earlier fixed-node
GFMC calculation \cite{,Boninsegni:1995_triang_GFMC} of the model gave a much
less accurate value for the order parameter $M$ as large as 62\% of
the classical value.

Finally, we note that the coupled cluster method (CCM), which will be
employed in the present study, has also been previously been applied
to the spin-$\frac{1}{2}$ triangular lattice HAFM
\cite{Zeng:1998_SqLatt_TrianLatt,Fa:2001_trian-kagome,SEKruger:2006_spinStiff,DJJFarnell:2014_archimedeanLatt}.
As will be explained in more detail in Sec.\ \ref{ccm_sec}, the CCM is
a size-extensive method that deals with infinite lattices ($N
\rightarrow \infty$) from the outset.  Nevertheless, results are
obtained at various levels of truncation in a well-defined and
systematic sequence of hierarchical approximations, namely the
lattice-animal-based subsystem (LSUB$m$) scheme described in Sec.\
\ref{ccm_sec}.  The only approximation then made is to extrapolate to
the exact limit $m \rightarrow \infty$ of the truncation index $m$.
The earliest CCM results
\cite{Zeng:1998_SqLatt_TrianLatt,Fa:2001_trian-kagome} were based on
relatively low-order approximations with $2 \leq m \leq 6$, and gave
an extrapolated result for the order parameter of the
spin-$\frac{1}{2}$ triangular lattice HAFM of $M \approx 51\%$ of the
classical value.  Later results based on more accurate extrapolations
with $2 \leq m \leq 8$ \cite{SEKruger:2006_spinStiff} and $4 \leq m
\leq 10$ \cite{DJJFarnell:2014_archimedeanLatt} gave much improved results of $M
\approx 42.7\%$ and $37.3\%$ of the classical value, respectively.
Both are in excellent agreement with the corresponding results using
the DMRG \cite{White:2007_triang_DMRG}, GFMC \cite{Capriotti:1999_trian}, ED
\cite{Richter:2004_triang_ED}, and SE \cite{Zheng:2006_triang_ED}
methods.

\begin{figure*}[!tbh]
\begin{center}
\mbox{
\subfigure[]{\includegraphics[height=3.5cm]{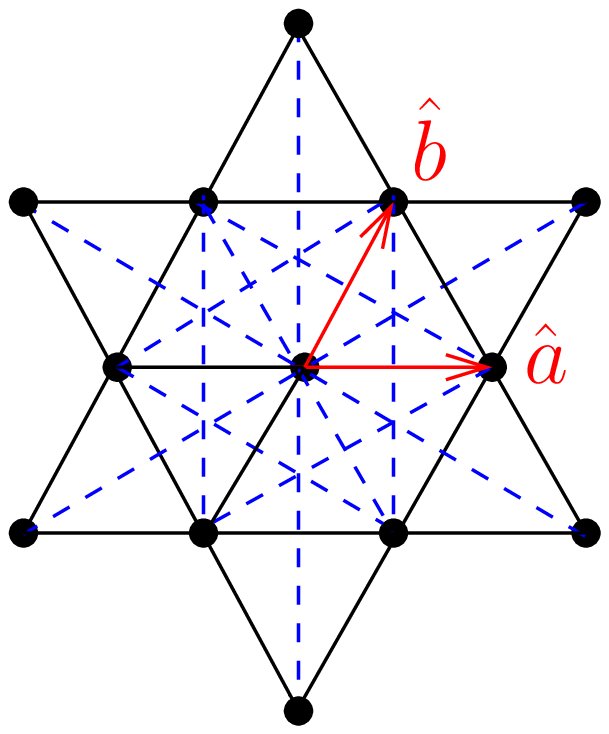}}
\quad
\subfigure[]{\includegraphics[height=3.5cm]{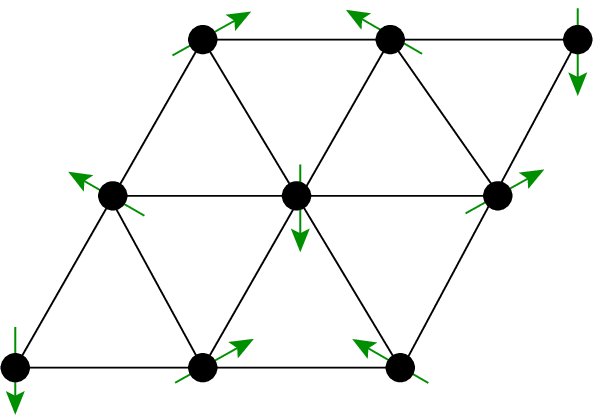}}
\quad
\subfigure[]{\includegraphics[height=3.5cm]{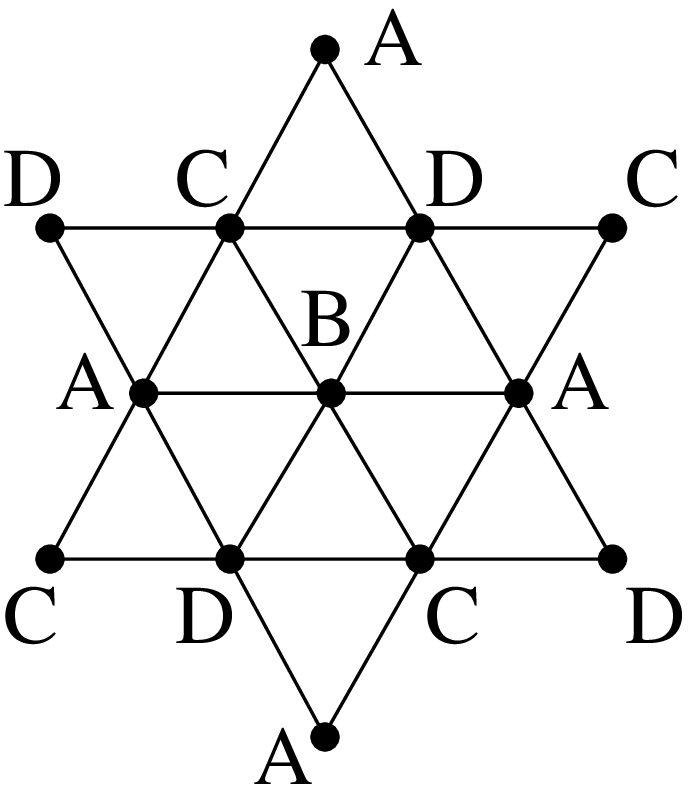}}
\quad
\subfigure[]{\includegraphics[height=3.5cm]{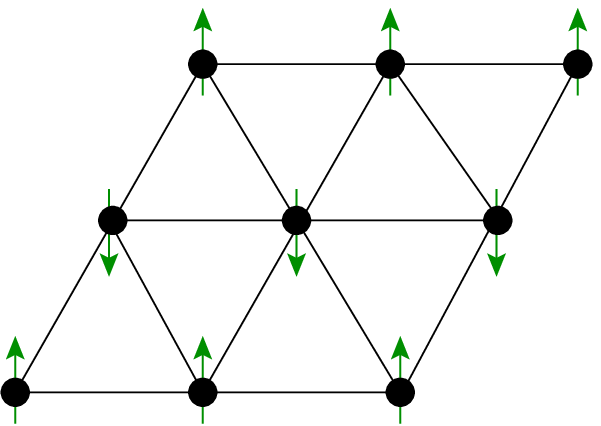}}
}
\caption{(Color online) The $J_{1}$--$J_{2}$ model on the triangular
  lattice with $J_{1}>0$ and $J_{2}>0$, showing (a) the bonds ($J_{1}
  \equiv$ -----~; $J_{2} \equiv \textcolor{blue}{- - -}$) and the
  Bravais lattice vectors $\hat{a}$ and $\hat{b}$; (b) the
  120$^{\circ}$ N\'{e}el antiferromagnetic (AFM) state; (c) the
  infinitely degenerate family of classical 4-sublattice ground states
  on which the spins on the same lattice are parallel to each other,
  with the sole constraint that the sum of four spins on different
  sublattices is zero; and (d) one of the three degenerate striped AFM
  states.  For the two states shown the arrows represent the
  directions of the spins located on lattice sites \textbullet.}
\label{model}
\end{center}
\end{figure*}  

Thus, by now, there is a rather clear consensus that the
spin-$\frac{1}{2}$ HAFM on the triangular lattice retains the
3-sublattice 120$^{\circ}$ N\'{e}el ordering of the classical ($s
\rightarrow \infty$) version of the model, albeit with a significant
decrease in the magnetic order parameter $M$ to a value of around $(40
\pm 2)\%$ of the classical value, due to quantum fluctuations.
Nevertheless, the model retains interest, both experimentally and
theoretically.  On the experimental side it is now believed that the
spin-$\frac{1}{2}$ HAFM on the triangular lattice can rather
accurately be realized in the compound Ba$_{3}$CoSb$_{2}$O$_{9}$
\cite{YShirata:2012_triang_experim,TSusuki:2013_experim}, in which the
magnetic Co$^{2+}$ ions form uniform triangular lattice layers.  In
this compound the effective magnetic moment of the Co$^{2+}$ ions,
which possess true spin $s=\frac{3}{2}$, can be well described by an
$s=\frac{1}{2}$ pseudospin at low enough temperatures $T$ (such that
$k_{B}T$ is much smaller than the spin-orbit coupling), where the
magnetic properties are determined by the lowest Kramers doublet.
Unlike in earlier possible realizations of spin-$\frac{1}{2}$
triangular lattice HAFMs, such as Cs$_{2}$CuCl$_{4}$
\cite{RColdea:2001_triang_experim} and Cs$_{2}$CuBr$_{4}$
\cite{,TOno:2003_triang_experim,NAFortune:2009_triang_experim}, in
which the triangular lattice is spatially distorted, and thus with an
exchange interaction that is spatially anisotropic, the triangular
lattice in Ba$_{3}$CoSb$_{2}$O$_{9}$ is expected to be regular.

On the theoretical side the spin-$\frac{1}{2}$ triangular lattice HAFM
retains specific interest as a starting-point to consider ways of
extending the model to investigate the stability of the classical
120$^{\circ}$ 3-sublattice N\'{e}el order against applied
perturbations.  We know, in particular, that exotic (nonclassical)
non-magnetically ordered states for spin-lattice systems tend to be
favored quantum-mechanically in situations for which the classical ($s
\rightarrow \infty$) counterpart has two or more different forms of GS
ordering that are degenerate in energy.  At the classical level Villain
{\it et al}. \cite{Villain:1977_ordByDisord,
  *Villain:1980_ordByDisord} showed how thermal fluctuations could
select, through the so-called {\it order by disorder} mechanism, a
specific form of order, which has softer excitation modes and hence,
for a given low energy, a larger density of states and a larger
entropy.

The commonest cause of such classical GS degeneracy is when competing
interactions are present.  One such example is the so-called
$J_{1}$--$J_{2}$ model on the triangular lattice in which the NN
interactions with exchange coupling strength $J_{1}>0$ now compete with
next-nearest-neighbor (NNN) interactions with exchange coupling
strength $J_{2} \equiv \kappa J_{1} > 0$.  We are thus led to the
study of a model in which both geometric and dynamic forms of frustration
are present simultaneously.  In the present paper we will consider
this model for spins with $s=\frac{1}{2}$.  Although initial interest
will focus on determining the critical value $\kappa^{c}_{1}$ of the
frustration parameter $\kappa$ at which the 120$^{\circ}$ N\'{e}el
order vanishes, our overall aim here is to study the entire ($T=0$) GS
phase diagram of the spin-$\frac{1}{2}$ $J_{1}$--$J_{2}$ model on the
triangular lattice in the range $0 \leq \kappa \leq 1$ of the
frustration parameter, in the case $J_{1}>0$.

Since the CCM has been shown to describe very accurately the limiting
case of the model when $\kappa=0$, as discussed above, we shall employ
it here also when $\kappa \neq 0$.  The plan of the rest of the paper
is as follows.  The model itself is first discussed in Sec.\
\ref{model_sec}, where we also discuss its classical ($s \rightarrow
\infty$) limit.  The main elements of the CCM are then reviewed in
Sec.\ \ref{ccm_sec}, before presenting our results in Sec.\
\ref{results_sec}.  We end with a summary and discussion in Sec.\
\ref{summary_sec}.

\section{THE MODEL}
\label{model_sec}
The Hamiltonian of the $J_{1}$--$J_{2}$ model on the triangular lattice is given by
\begin{equation}
H = J_{1}\sum_{\langle i,j \rangle} \mathbf{s}_{i}\cdot\mathbf{s}_{j} + J_{2}\sum_{\langle\langle i,k \rangle\rangle} 
\mathbf{s}_{i}\cdot\mathbf{s}_{k}\,,
\label{eq1}
\end{equation}
where index $i$ runs over all triangular lattice sites, and indices $j$ and $k$ run over all
NN and NNN sites to $i$, respectively, counting each bond once
and once only.  Each lattice site $i$ carries a particle with spin 
$s=\frac{1}{2}$ and a spin operator ${\bf s}_{i}=(s_{i}^{x},s_{i}^{y},s_{i}^{z})$. 
The lattice and exchange bonds are illustrated in Fig.\ \ref{model}(a).
We consider the case where both the NN and NNN bonds are
antiferromagnetic (i.e., $J_{1}>0$ and $J_{2} \equiv \kappa J_{1}>0$).
Henceforth, with no loss of generality, we set $J_{1} \equiv 1$ to set
the overall energy scale.

The classical ($s \rightarrow \infty$) version of the model has been
discussed in some detail by Jolicoeur {\it et
  al}. \cite{Joliceur:1990_J1J2-triang}.  They showed that the
3-sublattice 120$^{\circ}$ N\'{e}el antiferromagnetic (AFM) state
illustrated in Fig.\ \ref{model}(b) exists for $\kappa \leq
\kappa^{{\rm cl}}_{1} = \frac{1}{8}$.  This state thus has an energy
per spin, $E/N$, given by
\begin{equation}
\frac{E_{{\rm N\acute{e}el}}}{Ns^{2}} = 3(\kappa - \frac{1}{2})\,. \label{EperN}
\end{equation}
At $\kappa=\kappa^{{\rm cl}}_{1}$ the system then undergoes a
first-order phase transition into an infinitely degenerate family (IDF)
of 4-sublattice N\'{e}el ground states illustrated in Fig.\
\ref{model}(c), in which the only constraint is
$\mathbf{s}_{A}+\mathbf{s}_{B}+\mathbf{s}_{C}+\mathbf{s}_{D}=0$, where
$\mathbf{s}_{i}$ denotes the spin on each of the four sublattices,
$i=A,B,C,D$, as shown.  Each member of this IDF has an energy per spin
given by
\begin{equation}
\frac{E_{{\rm IDF}}}{Ns^{2}} = -\kappa - 1 \,.
\end{equation}
This IDF of 4-sublattice N\'{e}el states was shown to form the stable
GS phase for $\kappa^{{\rm cl}}_{1} \leq \kappa \leq \kappa^{{\rm
    cl}}_{2}=1$.  At $\kappa = \kappa^{{\rm cl}}_{2}$ the system then
undergoes a second-order phase transition to an incommensurate spiral phase with energy per spin given by
\begin{equation}
\frac{E_{{\rm spiral}}}{Ns^{2}}=-\frac{1}{2}(3\kappa+\frac{1}{\kappa})\,.
\end{equation}
This state persists for all values $\kappa > \kappa^{{\rm cl}}_{2}$.
In the limiting case $\kappa \rightarrow \infty$, when the NN
interactions no longer contribute, the three sublattices effectively
decouple and each of them again has a 120$^{\circ}$ N\'{e}el
configuration of spins.

The immediate question that arises is whether quantum fluctuations
will lift the degeneracy of the classical IDF of states in the regime
$\frac{1}{8}<\kappa < 1$, by the order by disorder mechanism.  Thus,
it is well known that any accidental degeneracy that occurs in systems
that have continuous degrees of freedom is usually removed by either
thermal or quantum fluctuations \cite{Villain:1977_ordByDisord,
  *Villain:1980_ordByDisord,Korshunov:1986_ordByDisord,Henley:1987_ordByDisord,Chubukov:1991_ordByDisord}.
Various authors
\cite{Joliceur:1990_J1J2-triang,AChubukov:1992_J1J2triang,SEKorshunov:1993_J1J2triang}
have applied lowest-order [i.e., to $O(1/s)$] SWT and shown that to
this order the 2-sublattice striped states, one of which is
illustrated in Fig.\ \ref{model}(d), are energetically preferred among
the IDF family.  Korshunov \cite{Korshunov:1986_ordByDisord} also
asserts that thermal fluctuations at the classical level favor the same
collinear striped ordering as do quantum fluctuations.  ED
calculations on finite clusters \cite{Lecheminant:1995_J1J2triang}
also led credence to this finding.  The striped AFM states have
ferromagnetic ordering along one direction [viz., the horizontal one
in Fig.\ \ref{model}(d)] and AFM ordering along the other two
principal directions of the triangular lattice.  They are thus
threefold-degenerate and break the rotational invariance of the
system.

Various authors
\cite{Joliceur:1990_J1J2-triang,AChubukov:1992_J1J2triang,SEKorshunov:1993_J1J2triang,Baskaran:1989_J1J2trian,Gazza:1993_J1J2triang,Deutscher:1993_J1J2triang,Ivanov:1993_XXZ_triang,Sindzingre:1995_J1J2triang,Lecheminant:1995_J1J2triang,Manuel:1999_J1J2triang,Mishmash:2013_J1J2triang,Kaneko:2014_J1J2triang}
have studied the spin-$\frac{1}{2}$ version of the $J_{1}$--$J_{2}$
model on the triangular lattice, using a number of approximate
methods, with little consensus to date concerning the $T=0$ GS phase
diagram.  On the other hand, to the best of our knowledge, no
high-order, systematically improvable method has yet been applied to
this spin-$\frac{1}{2}$ model.  It is our aim here to apply one such
technique, namely the CCM, to the model, in the regime $0 \leq \kappa
\leq 1$ of most interest.

\section{THE COUPLED CLUSTER METHOD}
\label{ccm_sec}
The CCM (see, e.g., Refs.\
\cite{Zeng:1998_SqLatt_TrianLatt,Bishop:1987_ccm,Arponen:1991_ccm,Bishop:1991_XXZ_PRB44,Bishop:1991_TheorChimActa_QMBT,Bishop:1998_QMBT_coll,Fa:2004_QM-coll,Bi:2008_PRB_J1xxzJ2xxz,Bishop:2012_honeyJ1-J2,RFB:2013_hcomb_SDVBC,Bishop:2014_honey_XY}
and references cited therein) is one of the most powerful and most
versatile techniques of modern quantum many-body theory.  Amongst
applications in a great variety of fields in physics and chemistry, it
has, in particular, been applied with considerable success to a large
number of spin-lattice problems in quantum magnetism (see, e.g.,
Refs.\ \cite{Zeng:1998_SqLatt_TrianLatt,Fa:2001_trian-kagome,SEKruger:2006_spinStiff,DJJFarnell:2014_archimedeanLatt,Bishop:1991_XXZ_PRB44,Fa:2004_QM-coll,Bi:2008_PRB_J1xxzJ2xxz,Bishop:2012_honeyJ1-J2,RFB:2013_hcomb_SDVBC,Bishop:2014_honey_XY,Kruger:2000_JJprime,Bishop:2000_XXZ,Darradi:2005_Shastry-Sutherland,Schm:2006_stackSqLatt,Darradi:2008_J1J2mod,Bi:2008_JPCM_J1J1primeJ2,Bi:2008_PRB_J1xxzJ2xxz,Darradi:2008_J1J2mod,Bi:2009_SqTriangle,Richter2010:J1J2mod_FM,Bishop:2010_UJack,Bishop:2010_KagomeSq,Reuther:2011_J1J2J3mod,DJJF:2011_honeycomb,Gotze:2011_kagome,Bishop:2012_checkerboard,Li:2012_honey_full,Li:2012_anisotropic_kagomeSq,RFB:2013_hcomb_SDVBC,Li:2013_chevron,Bishop:2013_crossStripe,Li:2014_honey_XXZ}
and references cited therein).  The CCM is especially suitable for the
study of frustrated magnetic systems for which the main alternative
techniques are often limited in their applicability.  For example,
quantum Monte Carlo (QMC) methods are usually severely restricted in such
cases by the well-known ``minus-sign problem''.  Similarly, the ED of
finite lattice clusters is limited in practice to such relatively
small clusters that it can be rather insensitive to the details of
some subtle forms of phase order that might be present.  By contrast
to almost all of the alternative methods that are capable of
systematic improvement within well-defined hierarchical approximation
schemes, the CCM provides both a size-consistent and size-extensive
technique, which gives results from the outset in the thermodynamic
(infinite-lattice, $N \rightarrow \infty$) limit, with no need,
therefore, for any finite-size scaling of the results.

We now briefly outline the CCM methodology to solve the GS
Schr\"{o}dinger ket- and bra-state equations,
\begin{equation}
H|\Psi\rangle=E|\Psi\rangle\,; \quad \langle\tilde{\Psi}|H = E\langle\tilde{\Psi}|\,.
\end{equation}
In order to describe quantitatively the quantum correlations present
in the exact GS phase under study, in the CCM one refers them to a
suitable, normalized model (or reference) state $|\Phi\rangle$.  This
state is required only to be a fiducial vector (or generalized vacuum
state) with respect to a suitable set of mutually commuting,
many-particle creation operators $C^{+}_{I}$.  Here, the index $I$ is
a set-index that defines a multiparticle configuration, and the
requirement is that the set of states $\{C^{+}_{I}|\Phi\rangle\}$
completely span the ket-state Hilbert space.

A key element of the CCM is that, unlike in the
configuration-interaction method, in which the decomposition of
$|\Psi\rangle$ is made linearly in this set, it is now made in an
exponentiated form.  Specifically, we have the parametrizations
\begin{equation}
|\Psi\rangle=e^{S}|\Phi\rangle\,; \quad \langle\tilde{\Psi}|=\langle\Phi|\tilde{S}e^{-S}\,,  \label{exp_para}
\end{equation}
where the two correlation operators $S$ and $\tilde{S}$ are formally
decomposed as follows,
\begin{equation}
S=\sum_{I\neq 0}{\cal S}_{I}C^{+}_{I}\,; \quad \tilde{S}=1+\sum_{I\neq 0}\tilde{{\cal S}}_{I}C^{-}_{I}\,,  \label{correrlation_oper}
\end{equation}
where we define $C^{+}_{0}\equiv 1$ to be the identity operator, and
$C^{-}_{I} \equiv (C^{+}_{I})^{\dagger}$.  For the case of spin-lattice problems
of the type considered here the set-index $I$ simply represents any
subset of the entire set of lattice sites (with possible repeats of
any site-indices), as discussed more fully below.  It denotes a multispin-flip configuration with
respect to the model state $|\Phi\rangle$, with
$C^{+}_{I}|\Phi\rangle$ representing the corresponding wave function
associated with this configuration of spins.  Hence, the operators
$\{C^{+}_{I}\}$ and $\{C^{-}_{I}\}$ are sets of mutually commuting
creation and destructor operators, respectively, defined with respect
to the state $|\Phi\rangle$ taken as a (generalized vacuum) reference
state.  They are hence chosen to obey the respective relations,
\begin{equation}
\langle\Phi|C^{+}_{I} = 0 =
C^{-}_{I}|\Phi\rangle\,, \quad \forall I \neq 0\,.
\end{equation}
By construction, therefore,
the states defined by Eqs.\ (\ref{exp_para}) and (\ref{correrlation_oper}) obey the normalization conditions
$\langle\tilde{\Psi}|\Psi\rangle = \langle{\Phi}|\Psi\rangle =
\langle{\Phi}|\Phi\rangle = 1$.  

In order both to treat each lattice site on an equal footing and to
make the computational implementation of the technique as universal as
possible, it is very convenient to make a passive rotation of the spin
on each lattice site in each model state $|\Phi\rangle$ so that in its
own local spin-coordinate frame it points along the negative $z$ axis,
which we henceforth denote as the downward direction.  Such passive
rotations are canonical transformations that leave the underlying
SU(2) commutation relations unchanged, and which, therefore, have no
physically observable consequences.  In these local spin coordinates,
which are clearly unique to each model state, every model state then
takes the universal form
$|\Phi\rangle=|\downarrow\downarrow\downarrow\cdots\downarrow\rangle$,
and the Hamiltonian has to be rewritten accordingly in these spin
coordinates.  Similarly, in these local spin-coordinate frames,
$C^{+}_{I}$ takes a universal form, $C^{+}_{I} \rightarrow
s^{+}_{l_{1}}s^{+}_{l_{2}}\cdots s^{+}_{l_{n}}$, a product of
single-spin raising operators, $s^{+}_{l} \equiv
s^{x}_{l}+is^{y}_{l}$.  The set-index $I \rightarrow
\{l_{1},l_{2},\cdots , l_{n};\; n=1,2,\cdots , 2sN\}$ thus simply
becomes a set of (possibly repeated) lattice site indices, where $N
(\rightarrow \infty$) is the total number of sites.  In the case of an
arbitrary spin quantum number $s$, a spin raising operator $s^{+}_{l}$
can be applied a maximum number of $2s$ times, on a given site $l$.
Hence, in any set-index $I$ included in the expansions of Eq.\
(\ref{correrlation_oper}) a given site $l$ may appear no more than $2s$
times.  Hence, in the present case where $s=\frac{1}{2}$, each site
index $l_{j}$ included in any single set index $I$ may appear no more
than once.

The (formally complete) set of GS multispin $c$-number correlation coefficients 
$\{{\cal S}_{I},{\tilde{\cal
    S}}_{I}\}$ is now determined by requiring that the GS energy expectation
functional,
\begin{equation}
\bar{H}=\bar{H}\{{\cal S}_{I},{\tilde{\cal S}_{I}}\}\equiv
\langle\Phi|\tilde{S}e^{-S}He^{S}|\Phi\rangle\,,  \label{GS_E_xpect_funct}
\end{equation}
is minimized with respect to each of the coefficients $\{{\cal S}_{I},{\tilde{\cal
    S}}_{I}\,; \forall I \neq 0\,.\}$.  From Eqs.\ (\ref{correrlation_oper}) and (\ref{GS_E_xpect_funct}) we thus obtain the coupled sets of equations,
\begin{equation}
\langle\Phi|C^{-}_{I}e^{-S}He^{S}|\Phi\rangle = 0\,, \quad \forall I \neq 0\,,  \label{ket_eq}
\end{equation}
by minimizing with respect to the parameter $\tilde{\cal S}_{I}$, and
\begin{equation}
\langle\Phi|\tilde{S}e^{-S}[H,C^{+}_{I}]e^{S}|\Phi\rangle\,, \quad \forall I \neq 0\,,  \label{bra_eq}
\end{equation}
by minimizing with respect to the parameter $S_{I}$.  Equation
(\ref{ket_eq}) takes the form of a coupled set of nonlinear
multinomial equations for the set of creations parameters $\{{\cal
  S}_{I}\}$.  Once solved the parameters $\{{\cal S}_{I}\}$ are used as input to
the coupled set of linear equations for the
set of destruction parameters $\{\tilde{\cal S}_{I}\}$, given by Eq.\ (\ref{bra_eq}).  Once Eq.\
(\ref{ket_eq}) has been satisfied, the value of $\bar{H}$ at the
minimum, which is simply the GS energy, may be expressed in the form
\begin{equation}
E=\langle\Phi|e^{-S}He^{S}|\Phi\rangle=\langle\Phi|He^{S}|\Phi\rangle\,.
\end{equation}
Equation (\ref{bra_eq}) for the destruction parameters $\{\tilde{\cal
  S}_{I}\}$ may then also be written in the equivalent form of a set
of generalized eigenvalue equations,
\begin{equation}
\langle\Phi|\tilde{S}(e^{-S}He^{S}-E)C^{+}_{I}|\Phi\rangle\,, \quad \forall I \neq 0\,,  \label{bra_eq_alternative}
\end{equation}

Another key feature of the CCM is that although the operator $S$ is
exponentiated, the actual equations (\ref{ket_eq}) and (\ref{bra_eq_alternative})
that we solve are automatically of finite order in the coefficients
$\{{\cal S}_{I}\}$, thereby obviating the need for any artificial
truncation.  The reason is that in Eqs.\ (\ref{ket_eq}) and
(\ref{bra_eq_alternative}) the operator $S$ only appears in the combination
$e^{-S}He^{S}$, a similarity transformation of the Hamiltonian.  This
form may be exactly and simply expanded as the well-known nested
commutator sum.  This otherwise infinite sum then terminates {\it
  exactly} with the double commutator term, firstly because all of the
terms comprising $S$ in its expansion of Eq.\
(\ref{correrlation_oper}) commute with one another and are just simple
products of single spin-raising operators as described above, and
secondly because of the basic SU(2) commutation relations (and see,
e.g., Refs.\
\cite{Zeng:1998_SqLatt_TrianLatt,Fa:2001_trian-kagome,Fa:2004_QM-coll}
for further details).  Exact such terminations equally apply for the
GS expectation values of all physical observable.  One such, in which
we will be interested, is the magnetic order parameter, defined to be
the average local on-site magnetization,
\begin{equation}
M \equiv
-\frac{1}{N}\langle\tilde{\Psi}|\sum_{l=1}^{N}s^{z}_{l}|\Psi\rangle=-\frac{1}{N}\langle\Phi|\tilde{S}\sum^{N}_{l=1}e^{-S}s^{z}_{l}e^{S}|\Phi\rangle\,,  \label{M_eq}
\end{equation}
where $s^{z}_{l}$ is defined now with respect to the chosen local
spin-coordinate frame on lattice site $l$, for the particular model
state $|\Phi\rangle$ being employed.  

Thus, for the reasons stated above, the {\it only} approximation made
in a practical implementation of the CCM is to truncate the set of
indices $\{I\}$ in the expansions of the correlation operators $S$ and
$\tilde{S}$ in Eq.\ (\ref{correrlation_oper}).  It is worth noting in
this context that it may be shown \cite{Bishop:1998_QMBT_coll} that
the CCM exactly obeys both the Goldstone linked cluster theorem (and
hence size extensivity) and the important Hellmann-Feynman theorem at
{\it any} such level of truncation.  We use here the well-tested
(lattice-animal-based subsystem) LSUB$m$ scheme
\cite{Zeng:1998_SqLatt_TrianLatt,Fa:2001_trian-kagome,SEKruger:2006_spinStiff,DJJFarnell:2014_archimedeanLatt,Bishop:1991_XXZ_PRB44,Fa:2004_QM-coll,Bi:2008_PRB_J1xxzJ2xxz,Bishop:2012_honeyJ1-J2,RFB:2013_hcomb_SDVBC,Bishop:2014_honey_XY,Kruger:2000_JJprime,Bishop:2000_XXZ,Darradi:2005_Shastry-Sutherland,Schm:2006_stackSqLatt,Bi:2008_JPCM_J1J1primeJ2,Bi:2008_PRB_J1xxzJ2xxz,Darradi:2008_J1J2mod,Bi:2009_SqTriangle,Richter2010:J1J2mod_FM,Bishop:2010_UJack,Bishop:2010_KagomeSq,Reuther:2011_J1J2J3mod,DJJF:2011_honeycomb,Gotze:2011_kagome,Bishop:2012_checkerboard,Li:2012_honey_full,Li:2012_anisotropic_kagomeSq,RFB:2013_hcomb_SDVBC,Li:2013_chevron,Bishop:2013_crossStripe,Li:2014_honey_XXZ}
in which, at the $m$th level of approximation, one retains all
multispin-flip configurations $I$ that are defined over no more than
$m$ contiguous lattice sites.  Any multispin-flip configuration or
cluster is defined to be contiguous if every site is NN to at least
one other in the cluster.  The number, $N_{f}$, of such distinct fundamental
configurations is reduced by fully exploiting the space- and point-group
symmetries, as well as any conservation laws, that pertain to both the
Hamiltonian and the model state being used.  Nevertheless, $N_{f}$
increases rapidly as the LSUB$m$ truncation index $m$ is increased,
and it becomes necessary at the higher orders to use massive
parallelization together with supercomputing resources
\cite{Zeng:1998_SqLatt_TrianLatt,ccm}.  In the present work we employ
both the 120$^{\circ}$ N\'{e}el and the collinear striped AFM states
shown in Figs.\ \ref{model}(b) and \ref{model}(d), respectively, as
CCM model states, and we have been able to perform LSUB$m$
calculations for all values $m \leq 10$ in both cases.  For example,
at the LSUB10 level, the number of fundamental configurations that we
employ is $N_{f}=271099$ for the striped AFM state and $N_{f}=1054841$
for the 120$^{\circ}$ N\'{e}el AFM state.

\begin{figure*}[!t]
\mbox{
\subfigure[]{\scalebox{0.31}{\includegraphics[angle=270]{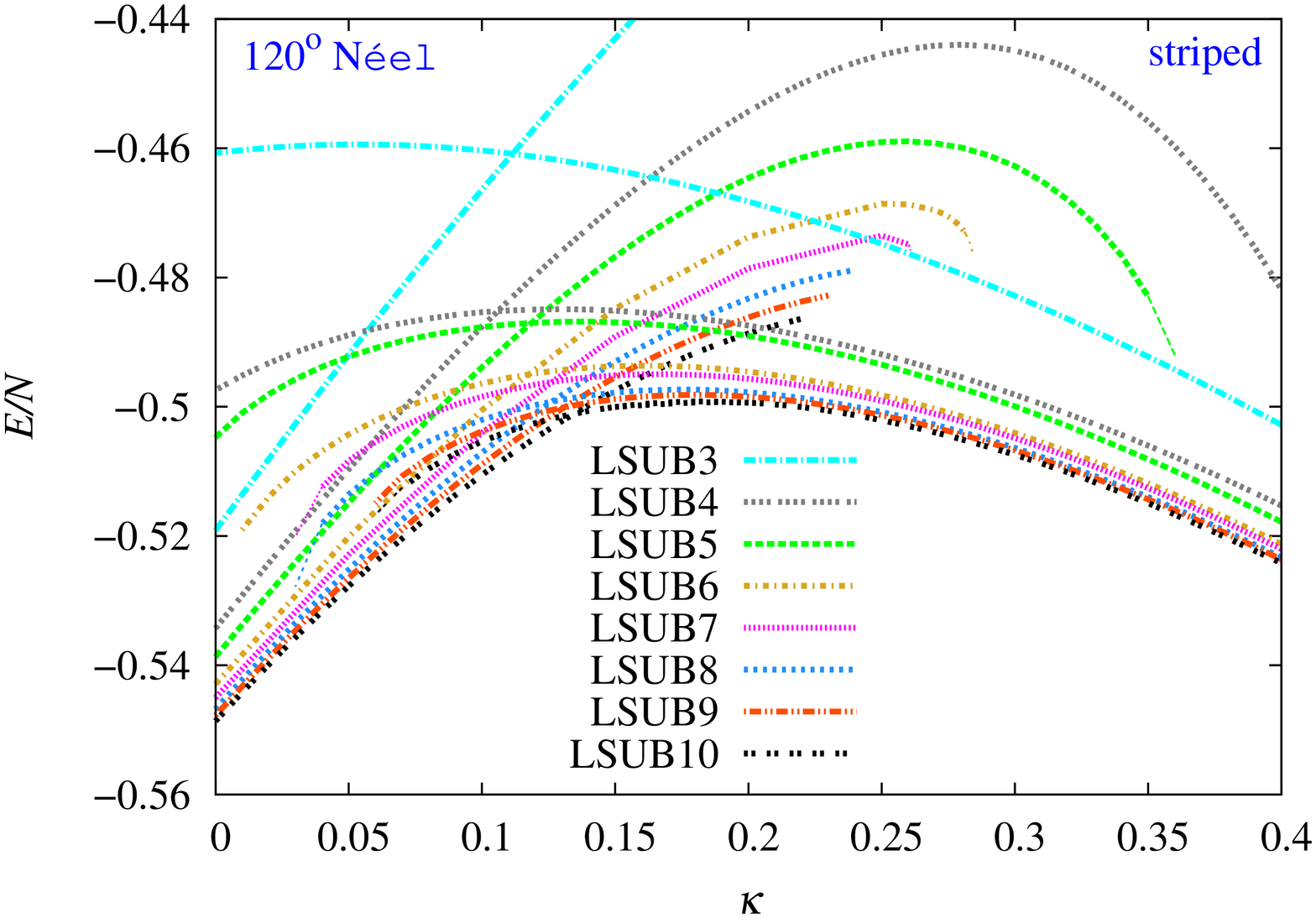}}}
\subfigure[]{\scalebox{0.31}{\includegraphics[angle=270]{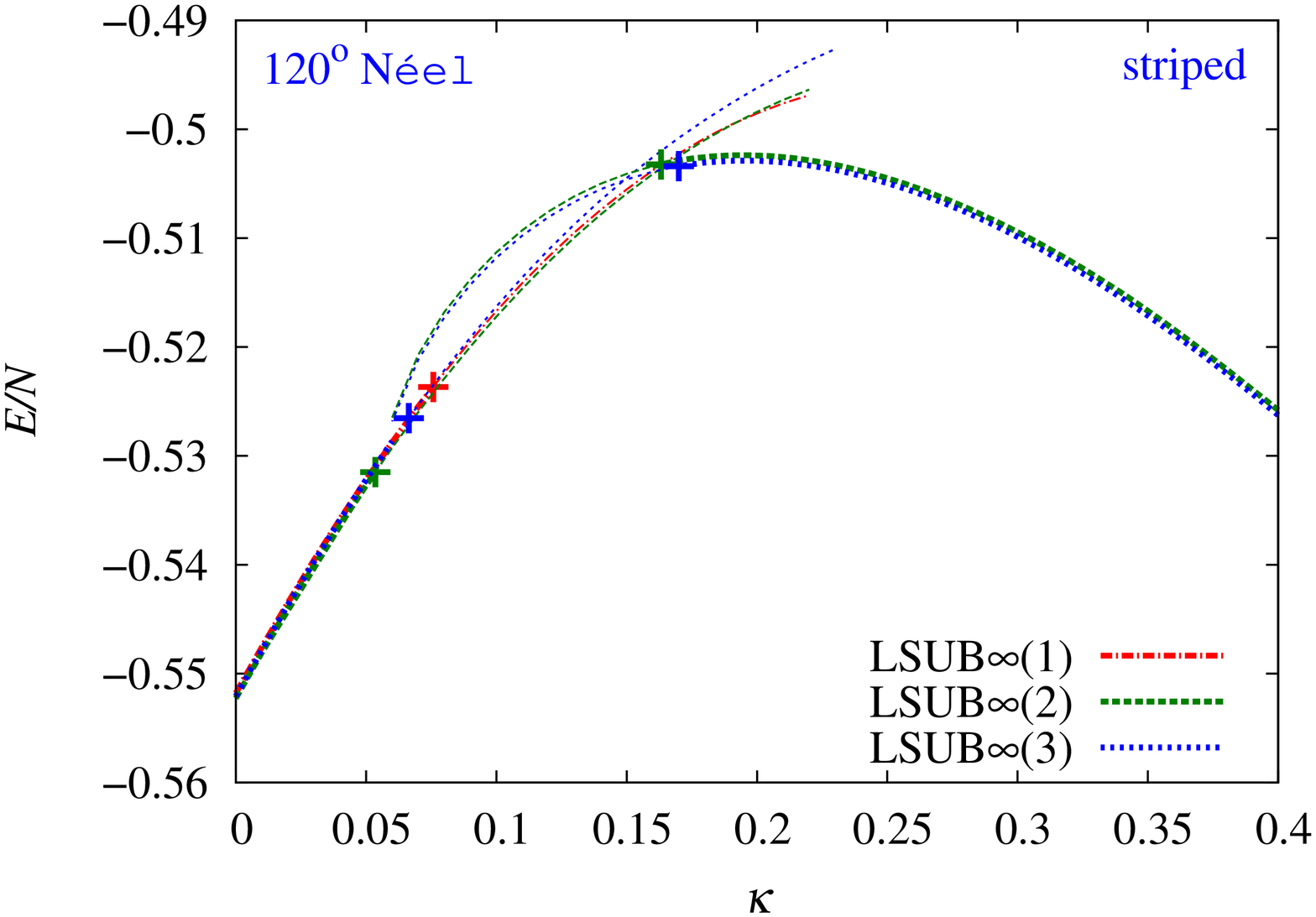}}}
}
\caption{(Color online) CCM results for the GS energy per spin, $E/N$,
  as a function of the frustration parameter $\kappa \equiv
  J_{2}/J_{1}$, for the spin-$\frac{1}{2}$ $J_{1}$--$J_{2}$ model on
  the triangular lattice (with $J_{1} \equiv +1$).  The left curves in each panel
  are based on the 120$^{\circ}$ N\'{e}el AFM state as CCM model
  state, while the right curves are similarly based on the striped AFM
  state as CCM model state.  (a) The LSUB$m$ results with $3 \leq m
  \leq 10$, shown out to their (approximately determined) termination
  points.  Portions of the curves with thinner lines denote the
  (approximately determined) unphysical regions where the magnetic
  order parameter takes negative values ($M<0$).  (b) The
  corresponding LSUB$\infty(k)$ extrapolations, based on Eq.\
  (\ref{E_extrapo}): the $k=1$ curve for the 120$^{\circ}$ N\'{e}el
  model state is based on LSUB$m$ results with $m=\{5,6,7,8,9,10\}$,
  while the $k=\{2,3\}$ curves for both model states are based on
  LSUB$m$ results with $m=\{4,6,8,10\}$ and $m=\{3,5,7,9\}$,
  respectively.  The plus (+) symbols mark those points where the
  corresponding extrapolated solutions have vanishing magnetic order
  parameters, $M=0$ [and see Fig.\ \ref{M}(b)].  Those sections of the
  curves beyond the plus (+) symbols, shown with thinner lines,
  indicate unphysical regions, where $M<0$ for these approximations
  (and see text for further details).}
\label{E}
\end{figure*}

Finally, the only extrapolation that we need to make is to the $m
\rightarrow \infty$ limit in the LSUB$m$ scheme, where our results for
all GS properties are, in principle, exact, since we make no other
approximations, and we work from the outset in the thermodynamic ($N
\rightarrow \infty$) limit.  The LSUB$m$ values for the GS energy per
spin, $E(m)/N$, converge very rapidly with increasing values of $m$.
We use the extrapolation scheme
\begin{equation}
E(m)/N = a_{0}+a_{1}m^{-2}+a_{2}m^{-4}\,,     \label{E_extrapo}
\end{equation}
which has been very widely tested and found to apply for a large variety of spin-lattice model 
\cite{Fa:2001_trian-kagome,DJJFarnell:2014_archimedeanLatt,Fa:2004_QM-coll,Bi:2008_PRB_J1xxzJ2xxz,Bishop:2012_honeyJ1-J2,RFB:2013_hcomb_SDVBC,Bishop:2014_honey_XY,Kruger:2000_JJprime,Bishop:2000_XXZ,Darradi:2005_Shastry-Sutherland,Schm:2006_stackSqLatt,Bi:2008_JPCM_J1J1primeJ2,Bi:2008_PRB_J1xxzJ2xxz,Darradi:2008_J1J2mod,Bi:2009_SqTriangle,Richter2010:J1J2mod_FM,Bishop:2010_UJack,Bishop:2010_KagomeSq,Reuther:2011_J1J2J3mod,DJJF:2011_honeycomb,Gotze:2011_kagome,Bishop:2012_checkerboard,Li:2012_honey_full,Li:2012_anisotropic_kagomeSq,RFB:2013_hcomb_SDVBC,Li:2013_chevron,Bishop:2013_crossStripe,Li:2014_honey_XXZ}.  As is to be expected, the expectation values of other GS quantities converge less rapidly than the energy.  For example, for most models studied previously that are either unfrustrated or contain only moderate amounts of frustration, the magnetic order parameter, $M$, defined in the local spin-coordinate frames by Eq.\ (\ref{M_eq}), typically follows a scheme with leading exponent $1/m$ 
\cite{Fa:2001_trian-kagome,Kruger:2000_JJprime,Bishop:2000_XXZ,Darradi:2005_Shastry-Sutherland,Bi:2009_SqTriangle,Bishop:2010_UJack,Bishop:2010_KagomeSq},
\begin{equation}
M(m) = b_{0}+b_{1}m^{-1}+b_{2}m^{-2}\,.   \label{M_extrapo_standard}
\end{equation}

On the other hand, for systems close to a QCP or when the magnetic
order parameter of the particular phase being studied is either zero
or close to zero, the extrapolation scheme of Eq.\
(\ref{M_extrapo_standard}) fits less well.  In such cases it typically
overestimates the amount of order present.  It usually also yields a
somewhat too large value of the critical strength of the frustration
interaction that is the primary driver for the corresponding phase
transition.  An alternative extrapolation scheme with leading exponent
$1/m^{1/2}$,
\begin{equation}
M(m) = c_{0}+c_{1}m^{-1/2}+c_{2}m^{-3/2}\,,   \label{M_extrapo_frustrated}
\end{equation}
has then been found both to provide an excellent fit to the LSUB$m$
results for a wide variety of models
\cite{Bi:2008_PRB_J1xxzJ2xxz,RFB:2013_hcomb_SDVBC,Bi:2008_JPCM_J1J1primeJ2,Richter2010:J1J2mod_FM,Reuther:2011_J1J2J3mod,DJJF:2011_honeycomb,Gotze:2011_kagome,Bishop:2012_checkerboard,Li:2012_honey_full,Li:2012_anisotropic_kagomeSq,RFB:2013_hcomb_SDVBC,Li:2013_chevron,Bishop:2013_crossStripe}
and also to yield more accurate values of the corresponding QCP.  In
practice, any of the extrapolation formulae of Eqs.\
(\ref{E_extrapo})--(\ref{M_extrapo_frustrated}), each of which
contains three fitting parameters, is ideally fitted to LSUB$m$
results with at least four different values of $m$, in order to obtain
accurate and robust fits.  In so far as there is no conflict with this
fitting rule, the lowest-order results with $m \leq 3$ are also
excluded, so far as practicable, since these results are usually rather
far from the asymptotic regime.

\section{RESULTS}
\label{results_sec}
We now present our CCM results for the spin-$\frac{1}{2}$
$J_{1}$--$J_{2}$ model (with $J_{1} \equiv 1$) on the triangular
lattice, using both the 120$^{\circ}$ N\'{e}el and the collinear
striped AFM states shown in Fig.\ \ref{model} as model states, and
employing the LSUB$m$ truncation scheme in each case for values of the
truncation index $m \leq 10$.  We first display the results for the GS
energy per spin, $E/N$, in Fig.\ \ref{E}.  Data are shown both for the
``raw'' LSUB$m$ results in Fig.\ \ref{E}(a) and for several
extrapolations based on Eq.\ (\ref{E_extrapo}) in Fig.\ \ref{E}(b),
using different LSUB$m$ data sets.

Several preliminary observations concerning the results shown are in
order.  Firstly, Fig.\ \ref{E}(a) clearly shows that the GS energy per
spin converges quite rapidly as a function of the LSUB$m$ truncation
index $m$ for both AFM phases based on the 120$^{\circ}$ N\'{e}el
state (left curves) and the striped state (right curves).  Secondly,
it is apparent from Fig.\ \ref{E}(a) that there is a marked even-odd
staggering effect for the raw LSUB$m$ results based on both model
states, which is particularly acute for the striped state.  In both
cases, however, the difference in the corresponding values for $E/N$,
at a given value of $J_{2}$, tends to be smaller between pairs of
LSUB$m$ results with $m=\{2n, 2n+1\}$ than between corresponding pairs
with $m=\{2n+1, 2n+2\}$, for integral values of $n$.  Thirdly, for a
given LSUB$m$ order of approximation, we see from Fig.\ \ref{E}(a)
that the corresponding pairs of curves for $E/N$, based on both model
states, cross one another at a value of the frustration parameter
$\kappa$ in the vicinity of the classical transition point at
$\kappa^{{\rm cl}}_{1}=\frac{1}{8}$.  Thus, there is clear preliminary
evidence of a quantum phase transition in the $s=\frac{1}{2}$ system
from a phase with 120$^{\circ}$ N\'{e}el AFM ordering at low values of
$\kappa$ to one with striped AFM ordering at high values of $\kappa$,
although it is not yet clear whether there is a direct transition
between these phases or whether it occurs via an intermediate state.
As we shall see below from a closer look at the extrapolated data, our
results are much more consistent with the latter scenario.

Fourthly, we note from Fig.\ \ref{E}(a) that both sets of curves,
based on each of the model states shown, display termination points at
specific values of $\kappa$.  In the case of the 120$^{\circ}$
N\'{e}el curves the termination points are upper ones, while for the
striped curves they are lower ones.  In each case the termination
points, which themselves depend on the LSUB$m$ truncation used, mark
the points beyond which there exist no real solutions to the
respective set of CCM equations, corresponding to Eq.\ (\ref{ket_eq}).
As is always the case, we see from Fig.\ \ref{E}(a) that as the
truncation index $m$ is increased, and the solution hence becomes more
accurate, the range of values of the frustration parameter $\kappa$
over which the corresponding LSUB$m$ approximations have real
solutions decreases.  Such CCM termination points have by now been
observed in many different spin-lattice problems, and are both well
documented and well understood (see, e.g., Refs.\
\cite{Fa:2004_QM-coll,Bishop:2010_UJack}), and as discussed further
below.  In particular, they provide a clear first signal of the
corresponding QCPs in the system under study, which denote the points
at which the respective forms of order shown by the model states
themselves melt.  In practice, however, what one finds is that
accurate solutions to the CCM LSUB$m$ equations of Eq.\ (\ref{ket_eq})
require an increasingly larger amount of computer power the nearer a
termination point is approached.  To obtain very accurate values of
the termination points themselves is thus computationally very costly.

A CCM LSUB$m$ termination point $\kappa^{t}_{i}(m)$ always arises at
the point where the solution with the $i$th model state to the
corresponding CCM equations given by Eq.\ (\ref{ket_eq}) become
complex.  Beyond such a point there actually exist two branches of
unphysical solutions, which are complex conjugates of one another.
Thus, in the region where the solution that tracks the true physical
solution is (necessarily) real, there actually exists another real
solution, which is numerically unstable and, hence, difficult to find
in practice.  The physical branch is, luckily, always the numerically
stable solution.  It is also always easily identifiable in practice as
the one that becomes exact in some known limit.  In all of our
displayed results, therefore, we display with confidence the branch
that represents the true (stable) ground state of the system.  This
physical branch then meets the corresponding unphysical branch at some
termination point (typically with infinite slope in curves such as
those in Fig.\ \ref{E}), beyond which no real solutions exist and the
two solutions branch into the complex plane as conjugate pairs.  As the
LSUB$m$ truncation index becomes larger, the two branches of real
solutions become closer, and as $m \rightarrow \infty$ they merge,
leaving the termination point as a mathematical branch point, which
represents the corresponding quantum critical point.  The LSUB$m$
termination points are thus themselves approximations to these
critical points.  Indeed, their $m \rightarrow \infty$ extrapolations
may be used as a method to estimate the position of the phase boundary
\cite{Fa:2004_QM-coll}.  Both since the LSUB$m$ termination points
themselves are computationally costly to obtain accurately, as already
noted, and also since we have other more accurate criteria at our
disposal to find the quantum critical points, as we shall see below,
we do not use this method here.

What is found here too, in common with many other applications of the
CCM to spin-lattice systems, and as we shall discuss more explicitly
below when we discuss our corresponding results for the magnetic order
parameter $M$, is that in the vicinity of the LSUB$m$ termination
points the respective solutions also become unphysical in the sense
that there exists a finite range of values of $\kappa$ for which $M$
becomes negative.  Thus, before the actual termination point of each
curve shown in Fig.\ \ref{E}(a), there exists a range of values of
$\kappa$ over which $M<0$.  Such (approximately determined) regions
where $M<0$ are shown in Fig.\ \ref{E}(a) by thinner lines than the
corresponding physical regions, where $M>0$, that are themselves
denoted by thicker lines.

It is perhaps worth emphasizing that the regions where $M<0$ occur
both for LSUB$m$ solutions with $m$ finite and for the corresponding
LSUB$\infty$ extrapolations.  We comment further on these points below
when the actual results for $M$ are discussed.

We show in Fig.\ \ref{E}(b) the corresponding extrapolated
(LSUB$\infty$) values, $a_{0}$, of the GS energy per spin, in each case
using Eq.\ (\ref{E_extrapo}) with various LSUB$m$ data sets.  In light
of the above-mentioned even-odd staggering effect in the raw LSUB$m$
results, we show separately extrapolations using the even-$m$ results
$m=\{4,6,8,10\}$ and odd-$m$ results $m=\{3,5,7,9\}$, for both model
states.  For the case of the 120$^{\circ}$ N\'{e}el AFM model state,
for which the staggering is not too pronounced, we also show the
extrapolated LSUB$\infty$ results $a_{0}$ from Eq.\ (\ref{E_extrapo})
using the data set $m=\{5,6,7,8,9,10\}$.  For both model states the
various extrapolated results are in excellent agreement with one
another.

We note that for both the even-$m$ and odd-$m$ extrapolations we have
tested explicitly for the applicability of Eq.\ (\ref{E_extrapo}).
Clearly, for any GS physical observable $X$, one may always check
directly for the correct leading exponent $\nu$ in the asymptotic ($m
\rightarrow \infty$) fitting formula,
\begin{equation}
X(m) = x_{0}+x_{1}m^{-\nu}\,,   \label{M_extrapo_nu}
\end{equation}
by fitting an LSUB$m$ set of results $\{X(m)\}$ to this form, and
treating each of the parameters $x_{0}$, $x_{1}$, and $\nu$ as fitting
parameters
\cite{RFB:2013_hcomb_SDVBC,Bishop:2014_honey_XY,RFB:2013_hcomb_SDVBC,Bishop:2000_XXZ,Darradi:2005_Shastry-Sutherland,Li:2013_chevron,Bishop:2013_crossStripe}.
For each of the even-$m$ and odd-$m$ data sets used in Fig.\
\ref{E}(b) the exponent $\nu$ from fitting to the form of Eq.\
(\ref{M_extrapo_nu}) is very close to the value 2, thereby justifying
the use of Eq.\ (\ref{E_extrapo}) in these cases.  While the even-odd
staggering effect leads to a fit with a comparatively worse value of
$\chi^{2}$ when both even and odd values of $m$ are used together than
those obtained using only even values or only odd values of $m$ separately,
such fits also yield values of $\nu$ close to 2 for the GS energy per spin.

\begin{figure*}[bth]
\mbox{
\subfigure[]{\scalebox{0.31}{\includegraphics[angle=270]{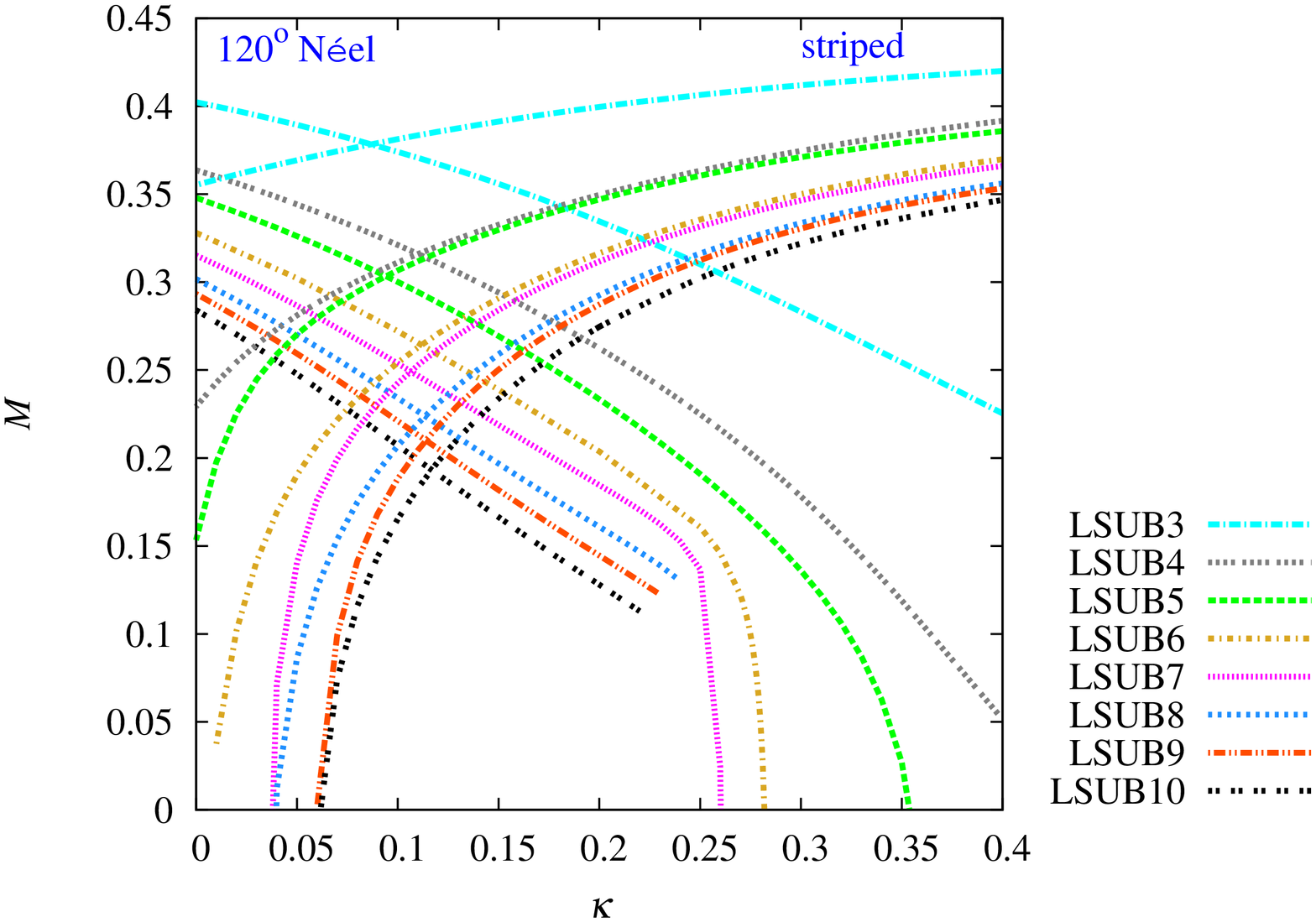}}}
\subfigure[]{\scalebox{0.31}{\includegraphics[angle=270]{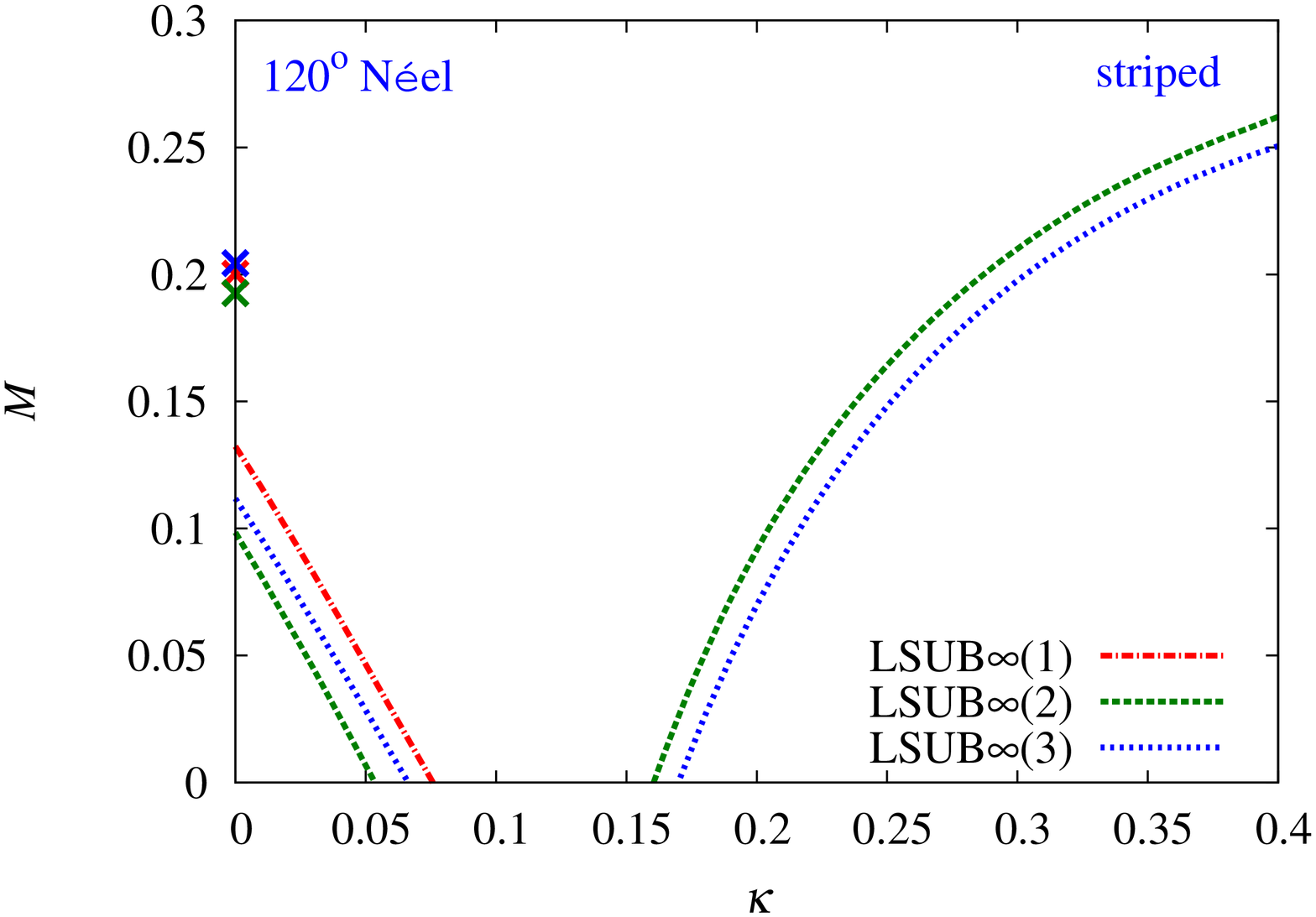}}}
}
\caption{(Color online) CCM results for the GS magnetic order, $M$, as
  a function of the frustration parameter $\kappa = J_{2}/J_{1}$, for
  the spin-$\frac{1}{2}$ $J_{1}$--$J_{2}$ model on the triangular
  lattice (with $J_{1}>0$).  The left curves in each panel are based
  on the 120$^{\circ}$ N\'{e}el AFM state as CCM model state, while
  the right curves are similarly based on the striped AFM state as CCM
  model state.  (a) The LSUB$m$ results with $3 \leq m \leq 10$, shown
  out to their (approximately determined) termination points.  (b) The
  corresponding LSUB$\infty(k)$ extrapolations, based on Eq.\
  (\ref{M_extrapo_frustrated}): the $k=1$ curve for the 120$^{\circ}$
  N\'{e}el model state is based on LSUB$m$ results with
  $m=\{5,6,7,8,9,10\}$, while the $k=\{2,3\}$ curves for both model
  states are based on LSUB$m$ results with $m=\{4,6,8,10\}$ and
  $m=\{3,5,7,9\}$, respectively.  As explained in the text the
  extrapolation scheme of Eq.\ (\ref{M_extrapo_frustrated}) is
  appropriate for the accurate determination of the QCPs at which $M
  \rightarrow 0$.  However, for zero or small dynamic frustration
  ($J_{2} \approx 0$) the scheme of Eq.\ (\ref{M_extrapo_standard})
  is appropriate.  Rather than crowd the figure with additional full
  curves based on Eq.\ (\ref{M_extrapo_standard}), we show with cross
  ($\times$) symbols the corresponding extrapolated values based on
  the 120$^{\circ}$ N\'{e}el state, using Eq.\
  (\ref{M_extrapo_standard}), for the case $\kappa=0$ only of the
  triangular-lattice HAFM.}
\label{M}
\end{figure*}

Before turning to our corresponding results for the magnetic order
parameter, $M$, it is worth discussing first the accuracy of our
results.  In order to do so let us consider the case $\kappa=0$ for
the pure triangular-lattice HAFM.  Thus, our extrapolated LSUB$\infty$
results using Eq.\ (\ref{E_extrapo}) are $E(\kappa=0)/N \approx
-0.55227 \pm 0.00011$ with LSUB$m$ results $m=\{4,6,8,10\}$;
$E(\kappa=0)/N \approx -0.55207 \pm 0.00001$ with LSUB$m$ results
$m=\{3,5,7,9\}$; and $E(\kappa=0)/N \approx -0.55180 \pm 0.00033$ with
LSUB$m$ results $m=\{5,6,7,8,9,10\}$, and where in each case the
errors quoted are solely those associated with the respective fits.  A
careful analysis of the errors yields our best estimate,
$E(\kappa=0)/N=-0.5521(2)$.  This may be compared, for example, with
the values $E(\kappa=0)/N=-0.5502(4)$ from a linked-cluster SE
technique \cite{Zheng:2006_triang_ED}; $E(\kappa=0)/N=-0.5415$
\cite{Bernu:1994_triang_ED} and $E(\kappa=0)/N=-0.5526$
\cite{Richter:2004_triang_ED} from two separate ED analyses of small
clusters of size $N \leq 36$
\cite{Bernu:1994_triang_ED,Richter:2004_triang_ED};
$E(\kappa=0)/N=-0.5458(1)$ from a GFMC technique
\cite{Capriotti:1999_trian}; $E(\kappa=0)/N=-0.5533$ from a Schwinger
boson mean-field theory (SBMFT) approach with $O(1/N)$ Gaussian
fluctuations included \cite{Manuel:1998_triang_SBMF}; and
$E(\kappa=0)/N=-0.5358$ and $E(\kappa=0)/N=-0.5468$ from leading-order
and second-order SWT
\cite{Miuyaki:1992_triang_SWT,Chernyshev:2009_triang_SWT},
respectively.  Our present value may also be compared with the value
$E(\kappa=0)/N=-0.5529$ from a recent CCM analysis
\cite{DJJFarnell:2014_archimedeanLatt} of the spin-$\frac{1}{2}$ HAFMs
on all 11 Archimedean lattices.  Although the raw LSUB$m$ results of
this latter work are identical with those obtained here for the
$\kappa=0$ case, the extrapolated value quoted there
\cite{DJJFarnell:2014_archimedeanLatt} is based on all results with $4
\leq m \leq 10$.  Due to the even-odd staggering effect present in
this case, we believe that the corresponding result
$E(\kappa=0)/N=-0.5529$ is skewed by including unequal number of even
and odd $m$ values in the fit.  Our own result, quoted above,
$E(\kappa=0)/N=-0.5521(2)$, is accordingly more accurate.  Finally, it
may be worth pointing out that, although, for example, the value
$E(\kappa=0)/N=-0.5533$ cited above from SBMFT lies below our result,
the SBMFT method is not variational and hence does not provide a
rigorous upper found to the GS energy.

Figure \ref{M} displays our corresponding results for the GS
magnetic order parameter, $M$, of Eq.\ (\ref{M_eq}) to those shown in
Fig.\ \ref{E} for the GS energy per spin, $E/N$.  
We see clearly from
Fig.\ \ref{M}(a) that at every LSUB$m$ level of approximation the
120$^{\circ}$ N\'{e}el AFM order vanishes at some upper critical value
$\kappa^{c}_{1}(m)$, whilst the striped AFM order vanishes at some
lower critical value $\kappa^{c}_{2}(m)$.  These are the respective
values used in Fig.\ \ref{E}(a) to demarcate the unphysical regions
whose $M<0$, shown by thinner lines, in the cases where this applies
to our data.  Once again, the even-odd staggering effect is clearly
visible in Fig.\ \ref{M}(a) for the results based on both model
states, particularly so for those based on the striped AFM state, for
which it is rather striking.

We note that it may not be obvious, a {\it priori}, why the LSUB$m$
regions with $M<0$ are necessarily unphysical.  Indeed, they could
simply arise because the quantization axes have been chosen
incorrectly.  However, what is found in practice is that the
corresponding LSUB$m$ critical values $\kappa^{c}_{i}(m)$ converge
relatively quickly as $m \rightarrow \infty$.  Furthermore, the extent
of the region between $\kappa^{c}_{i}(m)$ and the corresponding
LSUB$m$ termination point $\kappa^{t}_{i}(m)$, over which $M<0$,
shrinks as $m$ increases.  Finally, in the limit, $m \rightarrow
\infty$, $\kappa^{c}_{i}(\infty) = \kappa^{t}_{i}(\infty)$, and both
thus become equal to the corresponding quantum phase transition point.
In this sense, therefore, the unphysical regions in which $M<0$ are
artifacts of LSUB$m$ approximations with finite values of $m$.

The same LSUB$m$ data sets as were used in Fig.\ \ref{E}(b) for the GS
energy per spin extrapolations are also used in Fig.\ \ref{M}(b) for
the corresponding extrapolated curves for the GS magnetic order
parameter.  In all cases the curves have been obtained from the value
$c_{0}$ using the extrapolation scheme of Eq.\
(\ref{M_extrapo_frustrated}).  Once gain, we have checked explicitly,
by first fitting the LSUB$m$ values $M(m)$ of the GS magnetic order
parameter to Eq.\ (\ref{M_extrapo_nu}), that the extrapolation scheme
of Eq.\ (\ref{M_extrapo_frustrated}) is more appropriate than the
alternate scheme of Eq.\ (\ref{M_extrapo_standard}) for all of the
results based on the striped model state.  It is also the case for
most of the results based on the 120$^{\circ}$ N\'{e}el model state,
especially in the critical regime where $M \rightarrow 0$.  The only
exception is a very narrow region near $\kappa=0$, where the
extrapolation scheme of Eq.\ (\ref{M_extrapo_standard}) is clearly
preferred, as has been observed many time before, as discussed in
Sec.\ \ref{ccm_sec}.

The corresponding extrapolated values, $b_{0}$, obtained for the
spin-$\frac{1}{2}$ triangular-lattice HAFM (viz., at $\kappa=0$) are
shown in Fig.\ \ref{M}(b) by the cross symbols.  Once again, these
values may be used as benchmarks for comparison with those obtained by
other methods, and to discuss the overall quality of our CCM results.
Our extrapolated LSUB$\infty$ results using Eq.\
(\ref{M_extrapo_standard}) are $M(\kappa=0) \approx 0.193 \pm 0.002$
based on the data set $m=\{4,6,8,10\}$; $M(\kappa=0) \approx 0.204 \pm
0.003$ based $m=\{3,5,7,9\}$; and $M(\kappa=0) \approx 0.200 \pm
0.009$ based $m=\{5,6,7,8,9,10\}$.  In each case the quoted error is
that associated solely with the quality of the fit.  The even-odd
staggering effect is the cause of the larger error associated with the
fit using both even and odd values of $m$, compared to those
associated with the fits using only even or only odd values of $m$.  A
careful analysis of the errors yields our best estimate,
$M(\kappa=0)=0.198(5)$.

This value may again be compared with the corresponding value
$M(\kappa=0)=0.187$ from another recent CCM analysis
\cite{DJJFarnell:2014_archimedeanLatt} of the spin-$\frac{1}{2}$ HAFMs
on all 11 Archimedean lattices.  As discussed above for the GS energy
per spin results, although this latter work obtained raw LSUB$m$
results for the triangular-lattice HAFM that are identical to our own
$\kappa=0$ results, and although it also employed the extrapolation
scheme of Eq.\ (\ref{M_extrapo_standard}), the result quoted is based
on using the data set $m=\{4,5,6,7,8,9,10\}$.  Both the even-odd
staggering effect itself and the fact that the extrapolation is now
based on unequal numbers of even and odd $m$ values used in the fit,
now conspire to make the obtained value of 0.187 less accurate than
the value quoted here, 0.198(5).

Our value may again be compared with those from using the best of the
available alternate methods.  For example, a linked-cluster SE
analysis \cite{Zheng:2006_triang_ED} yields the value
$M(\kappa=0)=0.19(2)$; a recent ED analysis
\cite{Richter:2004_triang_ED} of small cluster of size $N \leq 36$
yields the $N \rightarrow \infty$ extrapolated value
$M(\kappa=0)=0.193$; a GFMC technique \cite{Capriotti:1999_trian}
based on clusters of size $N \leq 144$ yields the $N \rightarrow
\infty$ extrapolated value $M(\kappa=0)=0.205(10)$; and a DMRG
analysis \cite{White:2007_triang_DMRG} yields the $N \rightarrow
\infty$ extrapolated value $M(\kappa=0)=0.205(15)$.  All of these
modern values are seen to be in excellent agreement with one another,
with our own CCM result being now perhaps the most accurate available.
By contrast, the results from SWT and SBMFT are significantly larger.
Thus, the corresponding values from leading-order and second-order SWT
\cite{Miuyaki:1992_triang_SWT,Chernyshev:2009_triang_SWT} are
$M(\kappa=0)=0.2387$ and $M(\kappa=0)=0.2497$, respectively, while the
result from (lowest-order) SBMFT \cite{Gazza:1993_J1J2triang} is
$M(\kappa=0)=0.275$.

The LSUB$\infty$ extrapolated curves for $M$ shown in Fig.\ \ref{M}(b)
use the extrapolation scheme of Eq.\ (\ref{M_extrapo_frustrated}).
This scheme is particularly appropriate in the quantum critical
regimes, where $M$ becomes vanishingly small, as we have again
explicitly checked by first finding the leading exponent $\nu$ in fits
of $M$ to the scheme of Eq.\ (\ref{M_extrapo_nu}), using various LSUB$m$
data sets.  The corresponding values where $M \rightarrow 0$ are the
values shown in Fig.\ \ref{E}(b) on the extrapolated GS energy per
spin curves by the plus (+) symbols.  They provide our best estimates
for the QCPs, $\kappa^{c}_{1} \equiv \kappa^{c}_{1}(\infty)$ at which
the 120$^{\circ}$ N\'{e}el AFM order melts and $\kappa^{c}_{2} \equiv
\kappa^{c}_{2}(\infty)$ at which the striped AFM order melts.  We find
the explicit estimate $\kappa^{c}_{1} \approx 0.053$ from the
LSUB$\infty$ extrapolation scheme of Eq.\ (\ref{M_extrapo_frustrated})
using the LSUB$m$ data set $m=\{4,6,8,10\}$, with the corresponding
estimate $\kappa^{c}_{1} \approx 0.066$ from comparably using the
data set $m=\{3,5,7,9\}$.  The estimate obtained from using the data
set $m=\{5,6,7,8,9,10\}$ is $\kappa^{c}_{1} \approx 0.076$, although
the quality of this fit is considerably worse than those using only
even or only odd values of $m$, due to the even-odd staggering effect
discussed above, and hence this latter value comes with an appreciably
larger error.  The corresponding estimates obtained for
$\kappa^{c}_{2}$ are $\kappa^{c}_{2} \approx 0.163$ from using the
LSUB$m$ data set $m=\{4,6,8,10\}$, and $\kappa^{c}_{2} \approx 0.170$
from using the set of $m=\{3,5,7,9\}$.  On the basis of an analysis of
all our results our best estimates for the two QCPs are
$\kappa^{c}_{1} \approx 0.060(10)$ and $\kappa^{c}_{2} \approx
0.165(5)$.

In the concluding section we now summarize and discuss our results.

\section{SUMMARY AND DISCUSSION}
\label{summary_sec}
We have studied the spin-$\frac{1}{2}$ $J_{1}$--$J_{2}$ model on the
triangular lattice using the CCM in the case of AFM NN bonds
($J_{1}>0$) and AFM NNN bonds ($J_{2} \equiv \kappa J_{1} < 0$), in the
range $0 \leq \kappa \leq 1$ for the frustration parameter.  A big
advantage of the CCM is that, unlike most alternative accurate
methods, we work from the outset in the thermodynamic limit ($N
\rightarrow \infty$) of an infinite lattice, which hence obviates the
need for any finite-size scaling.

For the limiting case $\kappa=0$ of the triangular-lattice HAFM with
NN bonds only we find, in agreement with most other recent high-order
calculations, that the quantum $s=\frac{1}{2}$ model is magnetically
ordered, retaining the classical ($s \rightarrow \infty$)
120$^{\circ}$ N\'{e}el AFM order, albeit with a reduced value of the
GS magnetic order parameter $M$ (viz., the local on-site
magnetization), $M=0.198(5)$, compared to the classical value $M=0.5$.
In the same $\kappa=0$ limit the GS energy per spin is found to be
$E/N=-0.5521(2)$.  Both values are in excellent agreement with those
from other recent studies using high-accuracy methods, and both
possibly now represent the most accurate values available.

In the classical ($s \rightarrow \infty$) $J_{1}$--$J_{2}$ model on
the triangular lattice the 120$^{\circ}$ N\'{e}el AFM state is the
stable GS phase in the region $0 \leq \kappa \leq \kappa^{{\rm
    cl}}_{1}$, where $\kappa^{{\rm cl}}_{1} = \frac{1}{8}$.  At
$\kappa = \kappa^{{\rm cl}}_{1}$ there is then a first-order phase
transition to an IDF of 4-sublattice N\'{e}el states, which form the
stable GS in the region $\kappa^{{\rm cl}}_{1} < \kappa < \kappa^{{\rm
    cl}}_{2}$, where $\kappa^{{\rm cl}}_{2}=1$.  Lowest-order SWT and
ED calculations on finite clusters show that from this IDF of states,
the 2-sublattice striped states are energetically preferred over the
entire region $\frac{1}{8} < \kappa < 1$.  In the light of these
findings we have applied the CCM to the spin-$\frac{1}{2}$
$J_{1}$--$J_{2}$ model on the triangular lattice, using both the
3-sublattice 120$^{\circ}$ N\'{e}el and 2-sublattice striped AFM
states as model states.  

It is worth noting that, in principle, we could easily use other
candidate states from the IDF of 4-sublattice N\'{e}el states as CCM
model states, apart from the striped state used here.  By comparing
their extrapolated (LSUB$\infty$) GS energies we could then use the
CCM itself to provide evidence for a quantum order-by-disorder
selection of the striped state among the classical IDF.  However,
since many other methods provide strong evidence of the striped state
being selected, it seems somewhat redundant to do so.  Furthermore,
for the actual calculation of the QCP at $\kappa^{c}_{2}$, at which
quasiclassical ordering reappears for all $\kappa > \kappa^{c}_{2}$, it
  almost certainly suffices to use any of the IDF as a CCM model
  state.

Calculations have been performed in the
well-defined LSUB$m$ hierarchy of approximations, which becomes exact
in the limit $m \rightarrow \infty$.  High-order calculations have
been carried out for both quasiclassical states for values of the
truncation index $m \leq 10$, and we have discussed the extrapolations
to the $m \rightarrow \infty$ limit for both the GS energy per spin,
$E/N$, and the GS magnetic order parameter, $M$.

Our main finding is that the classical phase transition at
$\kappa^{{\rm cl}}_{1}=\frac{1}{8}$ is split, for the
spin-$\frac{1}{2}$ version of the model, into two quantum phase transitions at
$\kappa^{c}_{1} < \kappa^{{\rm cl}}_{1}$ and $\kappa^{c}_{2} >
\kappa^{{\rm cl}}_{1}$.  The quasiclassical 120$^{\circ}$ N\'{e}el AFM
order persists now only over the diminished range $(0 <)$ $\kappa <
\kappa^{c}_{1}$ under study, while the quasiclassical striped AFM
order persists over the (also diminished) range $\kappa^{c}_{2} <
\kappa$ ($< 1$) under study.  Our best estimates for the two
spin-$\frac{1}{2}$ QCPs are $\kappa^{c}_{1}=0.060(10)$ and
$\kappa^{c}_{2}=0.165(5)$

These findings may be compared with the corresponding results from
other methods.  For example, lowest-order (or linear) SWT
\cite{Ivanov:1993_XXZ_triang} predicts a quantum nonmagnetic phase for
the spin-$\frac{1}{2}$ case in the range $0.10 \lesssim \kappa
\lesssim 0.14$.  However, when leading-order, $O(1/s^{2})$,
corrections are included
\cite{AChubukov:1992_J1J2triang,Deutscher:1993_J1J2triang} this window
closes and the prediction then is that there is a direct first-order
transition at $\kappa \approx \frac{1}{8}$ between the two quasiclassical phases for the $s=\frac{1}{2}$
case, just as for the classical ($s \rightarrow \infty$) case.  By
contrast, lowest-order SBMFT \cite{Gazza:1993_J1J2triang} predicts a
first-order direct transition for the spin-$\frac{1}{2}$ model between
the two quasiclassical states at some critical value $\kappa^{c}
\approx 0.16$, with no interesting disordered phase.  At this critical
point the order parameter $M$ is reduced from its value
$M(\kappa=0)=0.275$, but is still nonzero, $M(\kappa \approx 0.16)
\approx 0.17$.  Now, however, when the leading-order corrections due
to Gaussian fluctuations are included \cite{Manuel:1999_J1J2triang} in
the SBMFT approach, there opens a window $0.12 \lesssim \kappa
\lesssim 0.19$, where the spin stiffness vanishes and the
quasiclassical 120$^{\circ}$ N\'{e}el and striped forms of magnetic
LRO both melt.  Clearly, the results of SWT and SBMFT approaches are
in conflict with one another.

Very recently, the phase diagram of the spin-$\frac{1}{2}$
$J_{1}$--$J_{2}$ model on the triangular lattice has been studied
using the variational Monte Carlo (VMC) method within various broad
classes of trial many-body wave functions
\cite{Mishmash:2013_J1J2triang,Kaneko:2014_J1J2triang}.  In a first
study \cite{Mishmash:2013_J1J2triang}, Mishmash {\it et al}. compared
the energies of the two quasiclassical AFM states, as modelled by
Jastrow-type wavefunctions, of the form pioneered by Huse and Elser
\cite{Huse:1988_triang_VMC}, which incorporate NN and NNN correlations
only, with that of a class of trial spin-liquid states with $d$-wave
symmetry.  On the basis of such a VMC calculation they find QCPs at
$\kappa^{c}_{1} \approx 0.05$ above which the 120$^{\circ}$ N\'{e}el
AFM order melts, and $\kappa^{c}_{2} \approx 0.18$ below which the
striped AFM order melts.  In between they find that a QSL state with
nodal $d$-wave symmetry has lower energy than either of the
surrounding quasiclassical states.  Clearly, the values so obtained
for the positions of the two QCPs are in good agreement with those
obtained in the present study.

Nevertheless, the Jastrow-type trial variational wavefunctions used by
Mishmash {\it el}.\ \cite{Mishmash:2013_J1J2triang} are relatively
inaccurate.  For example, for the case $\kappa=0$, Huse and Elser
\cite{Huse:1988_triang_VMC} obtained a VMC upper-bound value for the
GS energy per spin of $E(\kappa=0)/N \approx -0.5367$ with a trial
wavefunction (probably) containing more free parameters than that used
by Mishmash {\it et al}.\ for the 120$^{\circ}$ N\'{e}el AFM state, which is
appreciably above both our own value of $E(\kappa=0)/N=-0.5521(2)$ and
those from other accurate high-order methods quoted previously.
Although the difference in energy values may seem small, we note that
the classical value of the energy per spin is $-0.375$ from Eq.\
(\ref{EperN}).  Thus the best Jastrow-type wavefunction of Huse
and Elser, which included {\it all} two-spin interactions with a
spin-Jastrow factor proportional to $r^{-\sigma}_{ij}$, where $r_{ij}$
is the Euclidean distance between sites $i$ and $j$, together with
only the shortest-range three-spin term, still gives only about 92\%
of the nontrivial quantum part of the GS energy.  Since the energy differences
between competing phases are themselves small, as may be seen explicitly from Fig.\ \ref{E}(b), such errors may be highly
significant.

Indeed, the authors of a more recent VMC calculation
\cite{Kaneko:2014_J1J2triang} believe that the calculations of
Mishmash {\it et al}.\ may intrinsically overestimate the QSL phase,
due to the relative inaccuracy of their trial spin-Jastrow
wavefunctions for the quasiclassical AFM states.  Instead, Kaneko {\it
  et al}.\ \cite{Kaneko:2014_J1J2triang} calculate the ground and
low-lying excited states of the spin-$\frac{1}{2}$ $J_{1}$--$J_{2}$
model on the triangular lattice using a many-variable VMC approach.
They again find three locally stable states as candidates for the GS
phase.  These once more include the 120$^{\circ}$ N\'{e}el AFM state
and the striped AFM state, now together with a QSL state (with no LRO
order) of an unconventional critical (algebraic) type, characterized by
gapless excitations and a power-law decay of the spin-spin correlation
function.  Within their (enlarged) class of trial wavefunctions,
Kaneko {\it et al}.\ find that the 120$^{\circ}$ N\'{e}el AFM state is
favored for values $\kappa < \kappa^{c}_{1}=0.10(1)$, the striped AFM
state is favored for values $\kappa > \kappa^{c}_{2}=0.135(5)$, with
the critical QSL forming the stable GS phase for $\kappa^{c}_{1} <
\kappa < \kappa^{c}_{2}$.

Clearly, all variational studies are restricted by the class of trial
wavefunctions that they employ.  For that reason quantitative
estimates obtained from them for QCPs or phase boundaries must always
be treated with extreme caution.  What they can reveal, however, is
when certain states (e.g., of a QSL variety) become competitive
energetically with other more conventional (e.g., quasiclassical)
states.

One strength of the CCM used here is that it is certainly capable of
giving accurate values of QCPs and phase boundaries.  For that reason
we tend to believe that our own values for $\kappa^{c}_{1}$ and
$\kappa^{c}_{2}$ are intrinsically more accurate than those coming
from VMC calculations.  On the other hand, a weakness of the CCM as
implemented so far is that, despite giving accurate values of the QCPs
at which the two forms of quasiclassical order melt, our calculation
to date gives no information about the nature of the intermediate
state.  

Indeed, each of the CCM model states used here has been of the
independent spin-product type.  While it is certainly true that
solutions of the CCM are, to some extent, always tied to these
reference states, the relationship can be quite subtle.  For example,
it has been shown explicitly \cite{Farnell:2009_review} that exact
valence-bond crystal (VBC) states of the local dimer or plaquette
variety can also be described exactly within the CCM, starting from
the use of collinear independent-spin product states as model states.
More mundanely, one may also describe non-classical VBC ordering
within the CCM by the direct employment of valence-bond model states
\cite{Xian:1994_ccm_1Dchain} (e.g., on the square lattice, two- or
four-spin singlet product states) in place of the simpler single-spin
product states used here.  A complication of this approach, however,
is that a whole new matrix-operator formalism then needs to be created
for each new problem.  Both the Hamiltonian and the CCM bra- and
ket-state operators must then be rewritten in terms of the new matrix
algebra.  Once the commutation relations between the operators have
been found, the CCM equations must finally be derived and solved.
Although the whole procedure is formally straightforward, its
implementation in practice can be both tedious and computationally
intensive.  What is certainly much more difficult, however, is to use
directly a CCM model state of any of the usual QSL types.  Indeed, to
date, this has never been achieved.

\section*{ACKNOWLEDGMENT}

We thank the University of Minnesota Supercomputing Institute for the
grant of supercomputing facilities.

\bibliographystyle{apsrev4-1}
\bibliography{bib_general}

\end{document}